\documentclass[usenatbib]{mn2e}

\usepackage{epsf}

\newcommand{\lsun}{\ensuremath{\mbox{L}_{\odot}}}
\newcommand{\ghz}{\ensuremath{\mbox{GHz}}}

\newcommand{\kms}{\ensuremath{\mbox{km\,s}^{-1}}}
\newcommand{\mpc}{\ensuremath{\mbox{Mpc}}}
\newcommand{\wmhz}{\ensuremath{\mbox{W}\,\mbox{MHz}^{-1}}}
\newcommand{\mjy}{\ensuremath{\mbox{mJy}}}
\newcommand{\snu}{\ensuremath{S_{\nu}}}
\newcommand{\lnu}{\ensuremath{L_{\nu'}}}
\newcommand{\hsf}{\ensuremath{h_{65}}}
\newcommand{\lfir}{\ensuremath{L_{\rm FIR}}}
\newcommand{\loh}{\ensuremath{L_{\rm OH}}}

\title[Gigamasers and star-formation history]
      {Gigamasers: the key to the dust-obscured star-formation
       history of the Universe?}

\author[R.\,H.\,D.\ Townsend et al.]
       {R.\,H.\,D.\ Townsend,$^{1}$ R.\,J.\ Ivison,$^{2}$
        Ian Smail,$^{3}$ A.\,W.\ Blain$^{4}$ and D.\,T.\ Frayer$^{5}$\\
        $^1$ Department of Physics \& Astronomy, University College London, 
        Gower Street, London WC1E 6BT\\
        $^2$ Astronomy Technology Centre, Royal Observatory, Blackford
        Hill, Edinburgh EH9 3HJ\\
        $^3$ Department of Physics, University of Durham, South Road,
        Durham DH1 3LE\\
        $^4$ Institute of Astronomy, Madingley Road, Cambridge CB3 0HA\\
        $^5$ SIRTF Science Center, California Institute of Technology,
        MS 314-6, Pasadena, CA 91125, USA}

\pagerange{\pageref{firstpage}--\pageref{lastpage}}
\pubyear{2001}

\begin{document}

%% Title, abstract & keywords

\maketitle

\label{firstpage}

\begin{abstract}
We discuss the possibility of using OH and H$_2$O gigamasers to trace
the redshift distribution of luminous, dust-obscured, star-forming
galaxies. It has long been thought that ultraluminous, interacting
galaxies should host gigamasers due to their vast pumping infrared
(IR) luminosity, the large column density of molecules available to
populate the maser states and the turbulent motion of the gas in these
dynamically complex systems which allows unsaturated maser
emission. OH masers may thus be well-suited to the redshift-blind
detection of ultraluminous and hyperluminous infrared galaxies ($\lfir
\ge 10^{12}\ \lsun$) such as those uncovered by the SCUBA
submillimetre camera. The bandwidth requirement is low, $<1$\,GHz for
$z=1$--10 (lower still if additional redshift constraints are
available) and the dual-line 1665-/1667-MHz OH spectral signature can
act as a check on the reality of detections.
\end{abstract}


\section{Introduction} \label{sec:introduction}

The discovery of a distant population of luminous submillimetre
(submm) galaxies has revolutionised our understanding of galaxy
formation and evolution
\citep{Sma1997,Hug1998,Bar1998,Eal1999,Car2000}.  Debate continues
regarding the relative importance of obscured and unobscured star
formation, the fraction of active galactic nuclei (AGN) in the submm
galaxy population and the relationship, if any, between Lyman-break
galaxies and submm galaxies
\citep{AdeSte2000,Cha2000,Pea2000,Eal2000,Sma2002}; nevertheless, it
is clear that rest-frame far-infrared (far-IR) energy that has been
reprocessed by dust and redshifted into the waveband accessible to
SCUBA \citep{Hol1999} traces a galaxy population which makes a
significant, and possibly dominant, contribution to the star-formation
density at $z>1$.

It was realised from the outset that the most crucial piece of
information required to derive the history of obscured star formation
from the submm population is its redshift distribution, $N(z)$
\citep{Bla1999}. Knowledge of $N(z)$ breaks degeneracies in the models
and allows the nature of the galaxies to be explored, most importantly
providing estimates of their masses via observations of CO. Initial
efforts to target the first few SCUBA galaxies were very successful,
resulting in four redshifts from a sample of fifteen weakly lensed
galaxies in the \citet{Sma1998} sample \citep[][Frayer et al., in
prep]{Ivi1998,Sou1999,Ivi2000}, three of these have so far been
confirmed as massive and gas rich systems through CO line mapping
\citep[][Kneib et al., in prep]{Fra1998,Fra1999}.

However, the realisation that the majority of the submm population
have no plausible optical counterparts \citep[$I>26$,
e.g.,][]{Sma2000}, has meant that the possibility of a complete
optical spectroscopic survey has had to be dismissed
\citep[e.g.,][]{Bar1999}. Even in the IR, only around half of the
galaxies are identified by $K<22$, often as extremely red objects
\citep[EROs, $I-K>5$ --][]{Sma1999,Gea2000,Ivi2000}, giving little
hope to IR spectroscopists either.

As in other branches of extragalactic research, attention has
therefore been diverted to broadband photometric redshift techniques.
Unfortunately, the classical analysis of optical/IR photometry has
been shown to be misleading for even the few visible examples of these
very dusty galaxies, due to the complex effects of dust on the optical
spectral energy distribution, a problem which is further compounded by
the low photometric precision available for these faint galaxies.
However, \citet{CarYun1999} made a significant breakthrough with the
realisation that one of the strongest correlations in observational
astronomy -- between far-IR and radio emission \citep{Con1992} --
could be exploited to give an indication of redshift, based on the
submm fluxes and sensitive radio detections of the galaxies. Several
variants of this technique have been developed
\citep{CarYun2000,Bar2000,Dun2000}, though, based on comparison
against the few submm galaxies with known redshifts, none are
demonstrably better than the original. The technique has been used to
demonstrate convincingly that the median redshift of submm galaxies
must be in the range $2<z<3.5$, with no significant low-redshift tail
(Smail et al.\ 2000; cf.\ Smail et al.\ 1998; Lilly et al.\ 1999).

How else can progress be made towards the determination of the $N(z)$
for the submm population?  Some have proposed that searches for CO
rotational lines centred on submm galaxy positions are the way forward
\citep[e.g.,][]{Hug2000,Bla2000}. The CO lines are certainly expected
to be luminous, but the bandwidth requirements ($\sim 100\,\ghz$) of
such an approach are several orders of magnitude beyond the
capabilities of current instrumentation.  Moreover, prior to the
advent of ALMA \citep{Bro1999}, observers will be reliant on
current/planned 10--50m mm/submm single-dish facilities. As it has
long been known that even {\it interferometric} detections of CO
(where one benefits from stable baselines) in galaxies with {\it
known} redshifts are extraordinarily difficult \citep{Fra1998}; there
are obviously substantial technical difficulties to overcome when
undertaking redshift-blind, single-dish CO searches.  It is clear
therefore that we should explore other avenues to measure the
redshifts of the submm galaxy population and so understand more about
their detailed properties.

In this paper we propose an alternative method for determining the
redshifts of submm galaxies, based on the expectation that these
dusty, ultraluminous galaxies will exhibit similar behaviour of their
H$_{2}$O and OH maser activity to that observed in luminous IR
galaxies in the local Universe. We first discuss the background to
this proposal (\S2), based on the properties of H$_{2}$O and OH masers
in the local Universe, before investigating, in \S3, the feasibility
of these observations using current and future instrumentation.
Finally, in \S4, we state the main conclusions of this study.

\vspace*{-0.4cm}
\section{H$_{2}$O and OH Megamasers} \label{sec:masers}

Our proposed technique for determining the $N(z)$ of the submm
population is based on the expectation that luminous submm galaxies
(typically $\sim10^{12}\,\hsf^{-2}\,\lsun$, where $\hsf \equiv
H_{0}/65\,\kms\,\mpc^{-1}$) exhibit similar scaling of their H$_{2}$O
and OH maser activity to that observed in luminous IR galaxies in the
local Universe. Observations of OH emission in local luminous IR
galaxies, primarily due to the 1665-/1667-MHz ground rotational state
transitions, indicate a strong $\loh \propto \lfir^{2}$ correlation
between (isotropic) maser and far-IR luminosities
\citep[e.g.,][]{Baa1989}. This relationship has been explained in
terms of a model first proposed by \citet{Baa1985}, whereby an OH
population inversion, efficiently pumped by the far-IR flux, provides
unsaturated amplification of the background radio continuum; the
\citet{Con1992} far-IR/radio luminosity correlation then leads to the
observed quadratic dependence of $\loh$ on $\lfir$. More recent work
by \citet{Kan1996} has claimed that the correlation is merely a
consequence of Malmquist bias, the true relationship being closer to
$\loh \propto \lfir^{1.38}$, which \citet{Dia1999} have proposed to
result from the admixture of unsaturated and saturated emission.

Regardless of the precise character of any $\loh$/$\lfir$
relationship, however, it is clear
\citep[e.g.,][]{Baa1989,Baa1992b,Bri1998,DarGio2000} that powerful OH
masing is relatively common within the ultraluminous IR galaxy (ULIRG)
population, with upwards of 50 per cent of those galaxies with $\lfir
\ga 10^{11-12}\,\lsun$ supporting megamasers ($\loh \ga 10\,\lsun$) or
gigamasers ($\loh \ga 10^{3}\,\lsun$).  Based on the assumption that
the luminous submm galaxies are the high-redshift counterparts of
ULIRGs \citep{Sma2002}, it is therefore quite reasonable to expect
strong OH maser emission from a large proportion of the submm
population, which, if detectable, will allow the accurate
determination of their $N(z)$.

Although 22.235-GHz H$_{2}$O megamasers show somewhat different
characteristics to the OH maser systems
\citep[e.g.,][]{Bra1994,Bra1996}, similar arguments may be deployed in
favour of their potential as redshift markers for luminous submm
galaxies. H$_2$O masers are more typically associated with AGN and are
less useful tracers of star formation than their OH counterparts, but
a significant number of submm galaxies are known to harbour active
nuclei \citep[][Frayer et al., in prep]{Ivi1998,Ivi2000}. If detected,
high-resolution studies may yield information on their black hole
masses \citep{Miy1997}, in addition to their redshifts.  The bandwidth
requirement for detection of H$_{2}$O masers is only $\sim 9\ghz$
($\nu_{\rm obs}=2$--11\,GHz for $z=1$--10), similar to the bandwith
planned for the new Very Large Array (VLA) correlator. In the case of
OH masers this requirement drops below $1\,\ghz$ ($\nu_{\rm
obs}=165$--835\,MHz for $z=1$--10).  Since there is no significant
submm galaxy population at $z<1$ \citep{Sma2000}, the technique avoids
contamination by H\,{\sc i} emission from local galaxies
\citep{Bri1998} and we can thus search a relatively clean part of the
electromagnetic spectrum using existing interferometers with superb
instrumentation, high aperture efficiencies and large collecting areas
(e.g.\ VLA, Westerbork Synthesis Radio Telescope -- WSRT, Australia
Telescope Compact Array -- ATCA).

Inevitably, the difficulties with this new technique will lie in its
stringent sensitivity and dynamic-range requirements, as well as
radio-frequency interference in the 165--835-MHz (UHF) band (see,
e.g., {\tt http://www.atnf.csiro.au/SKA/intmit}). In the following
section, we estimate the sensitivity requirements by extrapolating
from previous observations of H$_{2}$O/OH megamasers and gigamasers in
luminous IR galaxies.

\vspace*{-0.4cm}
\section{Sensitivity requirements} \label{sec:sensitivity}

%
% FIGURE 1
%
\begin{figure}
\begin{center}
\leavevmode
\epsffile{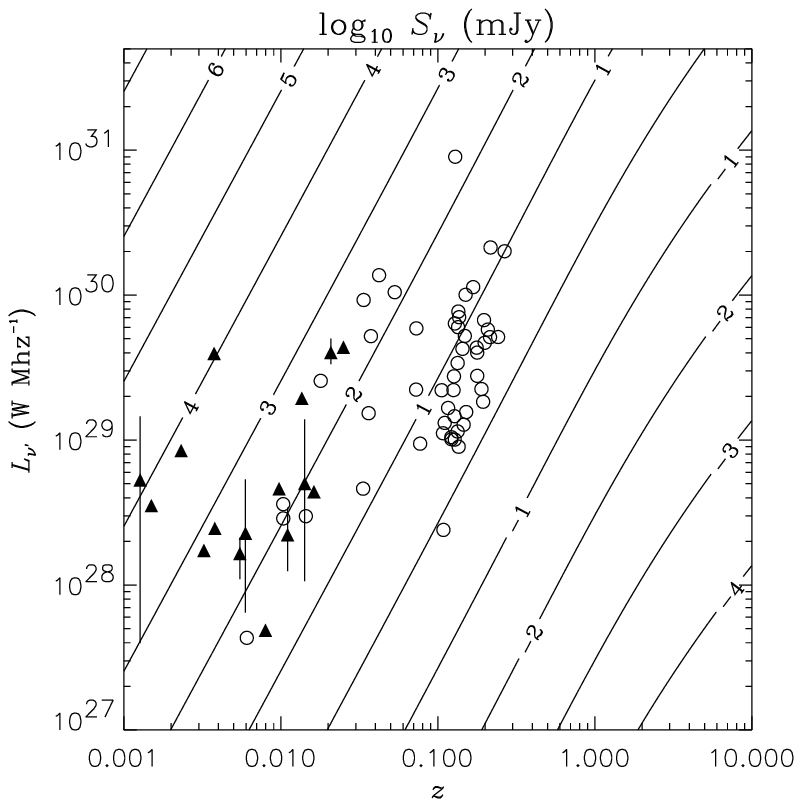}
\vspace*{-0.4cm}
\caption{Contour map of $\log_{10} \snu$, the peak observed flux
density, as a function of redshift, $z$, and peak rest-frame
luminosity density, $\lnu$. Also shown are observations of H$_{2}$O
(filled triangles) and OH (open squares) maser sources in luminous IR
galaxies; the data for these points have been calculated from
observations published by \citet{Baa1992a,Baa1992b}, \citet{Sta1992},
\citet{Bra1996} and \citet{DarGio2000,DarGio2001}. Where more than one
observation has been made of a given source, the vertical lines
indicate the spread of $\lnu$ about its average value.}
\label{fig:snu}
\end{center}
\end{figure}

Although the strength of extragalactic maser sources is often reported
in terms of the integrated line flux, we prefer to conduct our
estimations using the peak line flux density as it is this quantity
which determines whether a source is detectable with a given
instrument. For a maser at redshift $z$, emitting with a peak
rest-frame isotropic luminosity density of \lnu, the observed flux
density \snu\ at a frequency $\nu \equiv \nu'/(1+z)$ will be given by
\begin{equation}
\snu = (1+z) \frac{\lnu}{4\pi D_{L}^{2}(z)},
\end{equation}
where $D_{L}(z)$ is the luminosity distance
\citep[][]{Hog1999}. The $(1+z)$ factor in this expression accounts
for the linewidth narrowing due to the redshift, and partially offsets
the quadratic drop-off in \snu\ with distance.

In Fig.~\ref{fig:snu}, we plot a contour map of $\log_{10} \snu$ as a
function of $z$ and $\lnu$, where we have adopted $\hsf=1$,
$\Omega_{m}=0.3$ and $\Omega_{\Lambda}=0.7$ for the evaluation of
$D_L(z)$. Overplotted in the diagram are the loci of a selection of
the H$_{2}$O and OH maser sources observed in luminous IR galaxies
(see caption for references). Evidently, the H$_{2}$O sources appear
clustered at lower redshifts ($z \la 0.3$) and lower luminosity
densities ($\lnu \la 5\times 10^{29}\,\wmhz$); the OH masers tend to
be found at higher redshifts and luminosities. Whether this
distribution is intrinsic or due to selection effects remains unclear,
as does the apparent correlation between $z$ and $\lnu$, although this
is most likely a manifestation of Malmquist bias.

The three OH sources in the diagram with largest $\lnu$, correspond to
{\it IRAS}\,20100$-$4156 ($\lnu \sim 9.0\times10^{30}\,\wmhz$), {\it
IRAS}\,12032+0707 ($\lnu \sim 2.1 \times 10^{30}\,\wmhz$) and {\it
IRAS}\,14070+0525 ($\lnu \sim 2.0 \times 10^{30}\,\wmhz$). The latter,
discovered by \citet{Baa1992a}, is the most luminous gigamaser system
currently known (inferred $\loh \sim 1.05\times 10^{4}\,\lsun$);
however, this energy is distributed over a velocity width of $\sim
2,400\,\kms$, which explains why {\it IRAS}\,14070+0525 exhibits a
smaller $\lnu$ than the less luminous yet narrower-lined gigamasers in
{\it IRAS}\,12032+0707 \citep{DarGio2001} and {\it IRAS}\,20100$-$4156
\citep{Sta1989}. If these three sources were at redshift $z\sim 3$ (a
typical value anticipated for the submm galaxies) and left otherwise
unchanged, the observed flux densities would be $\snu \sim
0.089\,\mjy$, $\snu \sim 0.093\,\mjy$ and $\snu \sim 0.40\,\mjy$,
respectively. The paucity of points in Fig.~\ref{fig:snu} with $\snu <
1 \mjy$ illustrates that the detection of such faint sources is
probably beyond the capabilities of current technology.

However, the OH sources shown in Fig.~\ref{fig:snu} are all embedded
in {\it IRAS}-selected ULIRGs, and hence constitute a far-IR
flux-limited sample, with a redshift cutoff at $z\sim0.4$
\citep*{Cle1999}. The population \citep[e.g.,][]{Row2000} of
hyperluminous IR galaxies (HLIRGs), with $\lfir \ga 10^{13}\,\hsf^{-2}
\,\lsun$, suggests that more powerful OH masers may lie undetected at
redshifts $z\ga0.4$. The HLIRGs are especially promising candidate
hosts for OH masing: observations indicate that they are powered by
starburst activity \citep{Row2000} which may provide the turbulence
required for unsaturated masing to occur \citep{BurKom1990}. Recalling
that such unsaturated emission exhibits a quadratic $\loh \propto
\lfir^{2}$ behaviour, it is therefore possible that OH masers in
HLIRGs may exist with peak luminosity densities $\lnu$ approaching two
orders of magnitude greater than the values shown in
Fig.~\ref{fig:snu} for the ULIRGs. These immense luminosities would
render putative HLIRG gigamasers detectable out to $z\ga4$ at the
$\snu \sim 1\,\mjy$ level, close to the sensitivities of present-day
instrumentation, and within the grasp of facilities such as e-VLA,
e-MERLIN and --- ultimately --- the Square Kilometre Array (SKA);
however, issues relating to dynamic range and interference will need
to be addressed.

\vspace*{-0.3cm}
\section{Conclusions} \label{sec:conclusions}

OH and H$_2$O megamasers are common constituents of the most luminous
IR galaxies in the local Universe. The strong evolution in the
population of dusty starburst galaxies revealed by recent submm
observations \citep[e.g.,][]{Sma1997} should thus result in a
population of distant galaxies -- submm-selected galaxies (`SCUBA
galaxies') -- hosting extremely luminous masers. These lines should be
bright enough to be at the limit of detectability with current
instruments, but within the reach of e-VLA, e-MERLIN and ultimately SKA.

We propose that the redshifts of submm-selected galaxies, largely
beyond the reach of optical and IR spectroscopists, can be determined
using interferometric searches for these maser lines. Maser searches
have several clear advantages over other methods:
\vspace*{-0.2cm}
\begin{itemize}
\item
the bandwidth requirement is small, $<1$\,GHz for $z=1$--10 for OH
masers, smaller still if additional redshift constraints are available
(from their radio--submm spectral indices, for example \citealt{CarYun1999});
\item
the instantaneous survey area is limited by the primary beam of the
interferometer -- several degrees for an OH line search with e-VLA,
for example;
\item
interferometry permits some rejection of local radio-frequency
interference;
\item
the position of an emission line can be pinpointed accurately within
the primary beam, tying an emission line to a submm galaxy
unequivocally;
\item
the dual-line 1665-/1667-MHz OH spectral signature can act as an
important check on the line identification and the reality of
detections.
\end{itemize}

Armed with accurate redshifts for a significant proportion of the
submm galaxy population we could test the proposal that SCUBA galaxies
represent the massive progenitors of present-day ellipticals, using
measurement of their gas masses and fractions from interferometric CO
observations.  The redshift distribution for the SCUBA galaxies
derived from observations of megamasers would also remove the final
ambiguities in interpreting the contribution of this population to the
total star-formation density at redshifts of $z\sim 1$--5
\citep{Bla1999}.

\vspace*{-0.3cm}

\subsection*{Acknowledgements}

It is a pleasure to acknowledge valuable contributions from Chris
Carilli, Jeremy Yates, Padeli Papadopoulos, Phil Diamond, Rick Perley
and Paul van der Werf.  Furthermore, we thank the first referee for
the large amount of time they devoted to studying the paper. RHDT, IRS
and AWB acknowledge support from PPARC, the Royal Society and the
Raymond and Beverly Sackler Foundations.

\vspace*{-0.3cm}

%% Bibliography

\bibliographystyle{mn2e}
\bibliography{paper}

%% Document finish

\label{lastpage}

\end{document}